\documentclass[fleqn,usenatbib]{mnras}

\usepackage{newtxtext,newtxmath}

\usepackage[T1]{fontenc}
\usepackage{ae,aecompl}
\usepackage{times}

\usepackage{graphicx}	
\usepackage{amsmath}	
\usepackage{amssymb}	

\title[SIDM profile evolution]{The interplay of Self-Interacting Dark Matter and baryons in shaping the halo evolution}
\author[Giulia Despali et al.]{Giulia Despali$^{1}$\thanks{E-mail:gdespali@mpa-garching.mpg.de}, Martin Sparre$^{2,3,4}$, Simona Vegetti$^{1}$, Mark Vogelsberger$^{4}$, \newauthor
Jes\'us Zavala$^{5}$ and Federico Marinacci$^{4,6}$\\\\ 
$^{1}$ Max Planck Institute for Astrophysics, Karl-Schwarzschild-Strasse 1, 85748 Garching bei M{\"u}nchen, Germany\\
$^{2}$ Institut fur Physik und Astronomie, Universit{\"a}t Potsdam, Karl-Liebknecht-Str. 24/25, 14476 Golm, Germany\\
$^{3}$ Leibniz-Institut fur Astrophysik Potsdam (AIP), An der Sternwarte 16, 14482 Potsdam, Germany \\
$^{4}$ Department of Physics, Kavli Institute for Astrophysics and Space Research, Massachusetts Institute of Technology\\
$^{5}$ Center for Astrophysics and Cosmology, Science Institute, University of Iceland, Dunhagi 5, 107 Reykjavik, Iceland\\
$^{6}$ Institute for Theory and Computation, Harvard-Smithsonian Center for Astrophysics, 60 Garden Street, Cambridge, MA 02138, USA}

\pubyear{2018}

\begin{document}
\label{firstpage}
\pagerange{\pageref{firstpage}--\pageref{lastpage}}
\maketitle
\begin{abstract}
We use high-resolution hydrodynamical simulation to test the difference of halo properties in cold dark matter (CDM) and a self-interacting dark matter (SIDM) scenario with a constant cross-section of $\sigma^\text{T}/m_{\chi}=1\;\text{cm}^{2}\text{g}^{-1}$. We find that the interplay between dark matter self-interaction and baryonic physics induces a complex evolution of the halo properties, which depends on the halo mass and morphological type, as well as on the halo mass accretion history. While high mass haloes, selected as analogues of early-type galaxies, show cored profiles in the SIDM run, systems of intermediate mass and with a significant disk component can develop a profile that is similar or cuspier than in CDM. The final properties of SIDM haloes -- measured at $z=0.2$ -- correlate with the halo concentration and formation time, suggesting that the differences between different systems are due to the fact that we are observing the impact of self-interaction. We also search for signatures of self-interacting dark matter in the lensing signal of the main haloes and find hints of potential differences in the distribution of Einstein radii, which suggests that future wide-field survey might be able to distinguish between CDM and SIDM models on this basis. Finally, we find that the subhalo abundances are not altered in the adopted SIDM model with respect to CDM.

\end{abstract}




\section{Introduction}
One of the most important and long-lasting challenges of modern astrophysics is understanding and unveiling the nature of dark matter. Despite the many uncertainties on its composition, dark matter is considered the main driver of large-scale structure formation in the Universe. Moreover, the $\mathrm{\Lambda}$ Cold Dark Matter ($\mathrm{\Lambda}$CDM) model has been successful in explaining many aspects of galaxy formation and evolution, as well as in reproducing the density fluctuations in the early Universe with great accuracy \citep[e.g.][]{planck1_14}. On the other hand, N-body simulations based on cold dark matter (CDM) models present some discrepancies with observed quantities, such as the `missing satellites' , the `too-big-to-fail'  and the `core-cusp' problems \citep{klypin99,boylan-kolchin09,bullock17}.

Alternative models such as warm dark matter (WDM) or self-interacting dark matter (SIDM) have been proposed to solve these controversies. The main feature of WDM is the suppression in the abundance of small mass structures \citep{schneider92,lovell12,lovell14} and for this reason, it has been advocated as a solution in particular to the `missing satellites' problem \citep{lovell16}. On the other hand, even if the central densities of haloes are reduced in WDM, leading to a different concentration-mass relation \citep{ludlow16}, viable WDM models do not significantly alter the inner structure of haloes.
The original motivation to explore dark matter self-interaction was the fact that it is another possible source of suppression of low-mass (sub)haloes. However, the cross-section required for this to happen in (at least dark-matter-only) simulations is too large ($\simeq 10 \; \text{cm}^{2}\text{g}^{-1}$), above the most recent constraints \citep{zavala13,peter13b}. At the same time, one of the main signatures of SIDM lies in the creation of cores in the density profiles: these are generated by energy-exchange interactions, due to which the halo centre is heated and tends to become isothermal. The extent of the core and the central density depend on the SIDM cross-section: larger cross-sections produce larger cores \citep{yoshida00,dave01,colin02,rocha13,vogel12,vogel14b,vogel16}. If the self-scattering cross-section is $\simeq 1 \; \text{cm}^{2}\text{g}^{-1}$, at the scale of dwarf galaxies SIDM models can solve both the too-big-to-fail and the core-cusp problems \citep{zavala13}. Moreover, it has been shown that SIDM can explain the diverse shapes of galaxy rotation curves \citep{kamada17} and at the same time be consistent with observations of galaxy clusters.

However, it is important to point out that the majority of previous works make use of dark-matter-only simulations and thus ignore the interplay of self-interactions and baryonic physics. \citet{kaplinghat14} found through analytic calculations that for baryon-dominated systems, the thermalisation can significantly increase the SIDM central density, with the inner halo shape following the baryonic potential.
Recently, \citet{elbert18} and \citet{sameie18} used isolated simulation with an analytic galactic or disk potential in order to characterize the combined effect of SIDM and the main galaxy on the density profiles. They found that the cross-talk between dark matter and baryons leads to a variety of halo profiles, depending on the strength of the self-interaction and the galaxy properties; in particular, the presence of a compact stellar disk can lead to an increase of the overall central density through the core-collapse of the SIDM halo even beyond the CDM density \citep{koda11}. 

The aim of this work is to go one step further, by studying the baryonic effects not in an isolated and idealised scenario, but in full cosmological simulations to complement the efforts done in this regard in the past \citep{dicintio17,robles17,lovell18,harvey18}. 
Some recent works in this direction \citep{robertson18,robertson18b} concentrated on galaxy clusters, while here we focus on the mass range that goes from Milky-Way-like to massive elliptical galaxies. For this purpose we use high-resolution re-simulations of lens analogues from the Illustris simulation \citep{vogel14,vogel14c}, carried out in CDM and SIDM and with a realistic model for baryonic physics. We adopt a
 viable SIDM scenario with a constant cross-section of $\sigma_\text{T}/m_\chi= 1\; \text{cm}^{2}\text{g}^{-1}$ (model SIDM1 from \citealt{vogel14b}). We study the properties of simulated SIDM haloes, compare them with their CDM counterparts and then focus on the differences in the lensing signal from the two models.  

The paper is organised as follows: we describe the simulations in Section \ref{sec_sim} and in Section \ref{sec_main} we analyse the properties of the haloes as a whole and of the central galaxy.  In more detail, Section \ref{sec_prof} focuses on the properties of the sample at the final snapshot of our simulations ($z=0.2$), while in Section \ref{sec_evol} we discuss how these are related with the halo evolution history; Section \ref{sec_lensing} explores the impact of SIDM on lensing observables. In Section \ref{sec_sub} we study the subhalo population. Finally, we draw our conclusions in Section \ref{discussion}.

\section{Simulations} \label{sec_sim}

This section describes how we have created a sample of nine simulated galaxies. The selection was performed by \citet{despali17b} on galaxies from the large-scale cosmological simulation Illustris \citep{vogel14,Torrey14,Genel14}, which was carried out using the AREPO hydrodynamical code \citep{springel10}. Illustris reproduces observed quantities, such as the observed galaxy stellar mass function \citep{Torrey14}, the stellar mass -- halo mass relation \citep{Genel14}, and the star formation rate -- stellar mass relation \citep{Genel14,2015MNRAS.447.3548S}. 

\citet{despali17b} selected early type galaxies at $z\simeq 0.2$ to resemble strong gravitational lenses from the SLACS survey \citep{bolton06}, in terms of total mass, stellar mass, stellar effective radius and velocity dispersion. From this original selection, we identified a subsample of nine galaxies to resimulate for this work. 

The SLACS survey was optimised to spectroscopically identify strong gravitational lens-systems with bright early-type galaxies (ETGs) as deflectors. It has found nearly 100 lenses and lens candidates,  among which a sub-sample of 53 lenses have confirmed ETG morphology: these have well-determined central stellar velocity dispersion, stellar masses ranging from  $10^{10.5}$ to $10^{11.8}M_{\odot}$ \citep{auger09} and  estimated  total  masses   of  the  order  of  $10^{13}M_{\odot}$ \citep{auger10a}, as well as central projected masses determined from strong lensing. 

We have then re-simulated the subsample of nine galaxies with the zoom simulation method described in \citet{2016MNRAS.462.2418S}. In our zoom simulations the dark matter mass resolution is $4.4\times 10^6 M_{\odot}$, while the baryonic resolution is $9.1\times 10^5 M_{\odot}$. We use the IllustrisTNG physics model, which has a novel treatment of low-accretion rate black hole feedback \citep{2017MNRAS.465.3291W} and re-tuned galactic wind parameters \citep{pillepich18} in comparison to the original Illustris model (which is presented in \citealt{2013MNRAS.436.3031V}). The IllustrisTNG model was used in our re-simulations because it produces more realistic gas fractions and gas properties of massive galaxies and galaxy clusters than the original Illustris simulation model \citep{pillepich18,vogel18b}. As a consequence, some of the systems have a different morphology with respect to the original Illustris run, as it is detailed later on in the text. We make use of this increased variety in our sample generated by the new baryonic physics model: we investigate the effect of different galaxy morphologies on the dark matter distribution in the SIDM and CDM scenarios. Finally, we analyse the lensing properties of both early and late-types in our sample.

\begin{figure*}
\includegraphics[height=0.45\hsize]{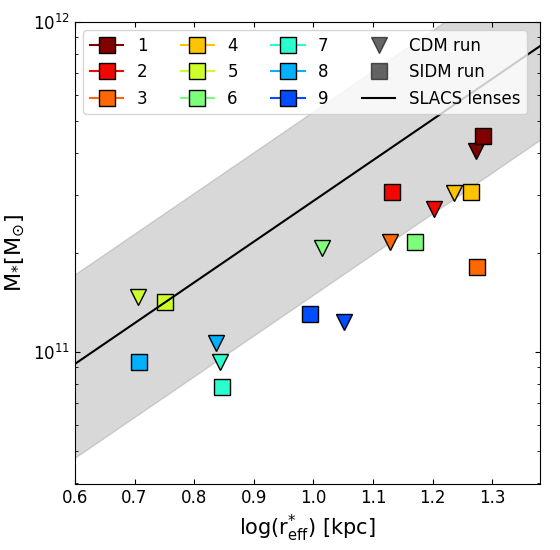}
\includegraphics[height=0.45\hsize]{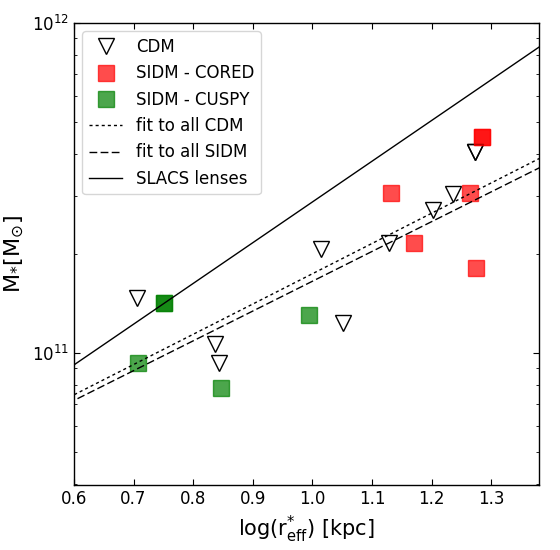}
\caption{Halo sample in the $r_\text{eff}^{*}$-$M_{*}$ plane at $z=0.2$. The black solid line shows the mean relation  derived for the SLACS lenses by \citet{auger10b}, together with the scatter of the data (gray band) calculated from the intrinsic scatter of the data and the uncertainty on the linear fit. \emph{Left panel}: the coloured squares and triangles show how the halo properties change between the CDM and SIDM run. \emph{Right panel}: we show the same points as in the left panel, but here the systems are color-coded according to the final profile properties; the  dotted and dashed black lines show the best-fit linear relation to the full sample of, respectively, CDM and SIDM haloes. \label{diagram}}
\end{figure*}

The AREPO code has been extended to include both elastic and multi-state inelastic SIDM~\citep{vogel12,vogel16,vogel18a}. Here we employ the elastic SIDM scheme with a constant cross-section. 
Specifically, each of our selected galaxies has been re-simulated both in a CDM and SIDM scenario with a constant cross-section of $\sigma_\text{T}/m_\chi= 1\; \text{cm}^{2}\text{g}^{-1}$ . The properties of the simulated halos are summarised in Table~\ref{sim1}.

\begin{table*} 
\begin{tabular}{cccccccc} 
\hline 
\multicolumn{8}{c}{CDM run} \\
ID & ID(original) & $M_\text{vir} [M_{\odot}]$ & $r_\text{vir}$ [kpc] & $M_{*} (\text{gal}) [M_{\odot}]$ & $r_\text{eff,*}$ [kpc] & D/T ($z=0.2$) & colour \\ 
\hline 
1 & 51 & 3.73 $10^{13}$ & 936 & 4.08 $10^{11}$ & 18.71 & 0.065 & \raisebox{-0.2\totalheight}{\includegraphics[width=0.01\textwidth]{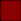}} \\
2 & 70 & 2.45 $10^{13}$ & 815 & 2.71 $10^{11}$ & 11.21 & 0.151&\raisebox{-0.2\totalheight}{\includegraphics[width=0.01\textwidth]{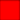}}\\
3 & 120 & 1.64 $10^{13}$ & 713 & 2.16 $10^{11}$ & 13.45 & 0.061&\raisebox{-0.2\totalheight}{\includegraphics[width=0.01\textwidth]{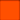}}\\
4 & 140 & 1.4 $10^{13}$ & 678 & 3.03 $10^{11}$ & 17.19 & 0.095&\raisebox{-0.2\totalheight}{\includegraphics[width=0.01\textwidth]{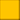}}\\
5 & 113 & 9.07 $10^{12}$ & 584 & 1.46 $10^{11}$ & 5.07 & 0.053&\raisebox{-0.2\totalheight}{\includegraphics[width=0.01\textwidth]{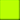}}\\
6 & 269 & 7.61 $10^{12}$ & 551 & 2.06 $10^{11}$ & 10.33 & 0.085& \raisebox{-0.2\totalheight}{\includegraphics[width=0.01\textwidth]{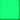}}\\
\hline
7 & 496 & 4.60 $10^{12}$ & 466 & 9.33 $10^{10}$ & 6.97 & 0.278 &\raisebox{-0.2\totalheight}{\includegraphics[width=0.01\textwidth]{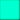}}\\
8 & 486 & 4.26 $10^{12}$ & 455 & 1.06 $10^{11}$ & 6.86 & 0.240 &\raisebox{-0.2\totalheight}{\includegraphics[width=0.01\textwidth]{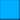}}\\
9 & 388 & 3.00 $10^{12}$ & 403 & 1.74 $10^{11}$ & 13.32 & 0.409& \raisebox{-0.2\totalheight}{\includegraphics[width=0.01\textwidth]{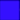}} \\
\end{tabular}
\begin{tabular}{cccccccc} 
\hline
\hline
\multicolumn{8}{c}{SIDM run} \\
ID & ID(original) & $M_\text{vir} [M_{\odot}]$ & $r_\text{vir}$ [kpc] & $M_{*} (\text{gal}) [M_{\odot}]$ & $r_\text{eff,*}$ [kpc] & D/T  ($z=0.2$) & color \\
\hline 
1 & 51 & 3.76 $10^{13}$ & 939 & 4.50 $10^{11}$ & 19.26 & 0.084 & \raisebox{-0.2\totalheight}{\includegraphics[width=0.01\textwidth]{figures/1}} \\
2 & 70 & 2.46 $10^{13}$ & 813 & 3.02 $10^{11}$ & 13.5 & 0.192 & \raisebox{-0.2\totalheight}{\includegraphics[width=0.01\textwidth]{figures/2}} \\
3 & 120 & 1.65 $10^{13}$ & 713 & 1.75 $10^{11}$ & 18.78 & 0.054 & \raisebox{-0.2\totalheight}{\includegraphics[width=0.01\textwidth]{figures/3}} \\
4 & 140 & 1.41 $10^{13}$ & 676 & 3.04 $10^{11}$ & 18.38 & 0.107 & \raisebox{-0.2\totalheight}{\includegraphics[width=0.01\textwidth]{figures/4}}  \\
5 & 113 & 8.64 $10^{12}$ & 574 & 1.42 $10^{11}$ & 5.64 & 0.116 & \raisebox{-0.2\totalheight}{\includegraphics[width=0.01\textwidth]{figures/5}} \\
6 & 269 & 7.51 $10^{12}$ & 548 & 2.14 $10^{11}$ & 14.40 & 0.096 & \raisebox{-0.2\totalheight}{\includegraphics[width=0.01\textwidth]{figures/6}} \\
\hline
7 & 496 & 4.42 $10^{12}$ & 459 & 7.81 $10^{10}$ & 7.02 & 0.284 & \raisebox{-0.2\totalheight}{\includegraphics[width=0.01\textwidth]{figures/7}} \\
8 & 486 & 3.95 $10^{12}$ & 441 & 9.31 $10^{10}$ & 5.10 & 0.099 & \raisebox{-0.2\totalheight}{\includegraphics[width=0.01\textwidth]{figures/8}} \\
9 & 388 & 2.98 $10^{12}$ & 403 & 1.85 $10^{11}$ & 11.66 & 0.342  & \raisebox{-0.2\totalheight}{\includegraphics[width=0.01\textwidth]{figures/9}} \\
\hline
\end{tabular}

\caption{Summary of halo properties: total halo mass $M_\text{vir}$, virial radius $r_\text{vir}$, stellar mass of the central galaxy $M_{*}$, stellar effective radius and disk/total mass ratio (D/T). The last column shows the colour used to represent each halo throughout the paper (in the plots where different systems are shown together); the colours are the same for corresponding haloes in the CDM and SIDM runs. The systems are ranked by total halo mass ($M_\text{vir}$); the last three systems have a developed disk in the CDM run (compact for systems 7 and 8, more extended and massive for system 9), while the others are classified as ETG-analogues.  \label{sim1}}

\end{table*}

\section{Properties of the main galaxy} \label{sec_main}

In this section we first analyse our sample at redshift $z=0.2$ (Sec. \ref{sec_prof}) and then explore the connection between the final halo properties and the halo evolution history (Sec. \ref{sec_evol}). Finally, we show the impact of SIDM on the distribution of the lensing observables (Sec. \ref{sec_lensing}).

\subsection{Mass distribution at z=0.2} \label{sec_prof}

Table \ref{sim1} summarizes the main properties of the nine simulated haloes at redshift $z=0.2$. The systems are ranked by total halo mass $M_{\rm vir}$
\citep[i.e. the mass within the radius that encloses a virial overdensity at the redshift considered, defined following][]{bryan98}. In general, the total halo mass differs only by a few percents between the two runs and the total stellar mass of the central galaxies is also very similar. The disk mass and thus the disk to total mass ratios (D/T) were calculated using the stellar circularities, as in previous works \citep{2009MNRAS.396..696S,2014MNRAS.437.1750M,teklu15,snyder15,2017MNRAS.467..179G} and in the same way that led to the selection from \citet{despali17b}. In more details, the classification is based on the specific angular momentum through  the parameter $\epsilon=J_{z}/J(E)$, where $J_{z}$ is  the specific angular momentum  and $J(E)$ is the maximum local angular momentum  of the stellar particles: the fraction of stars with $\epsilon>0.7$ is a common definition of the disk mass and we adopt it for this work. Halo 9 was explicitly selected to host an extended and massive disk, and thus it is by construction not an analogue of the SLACS lenses. On the other hand, Halo 7 and Halo 8 had a disk mass component below 20 per cent in the original Illustris run, but they develop a more massive disk in the new CDM run with the IllustrisTNG model and so we classify them as disk-galaxies in this work. In the SIDM run, the disk component of system 8 at the final time is much smaller, while the other two systems maintain a similar morphology as in the corresponding CDM run. From this point of view, Systems 1 to 6 can instead be classified as early-type galaxies in both runs. 

The left panel of Figure \ref{diagram} shows the position of our systems in the $r^{*}_\text{eff}-M_{*}$ plane. Each system corresponds to a different colour - this colour scheme is maintained throughout the whole paper when the haloes are shown together (see Table \ref{sim1}). 
In general, the IllustrisTNG model produces lower stellar masses \citep[at fixed halo mass - ][]{pillepich18} and smaller sizes \citep[at fixed stellar mass - ][]{genel18}: this explains why the systems from the CDM run (coloured triangles) tend to lie below the mean relation for the SLACS lenses, while the corresponding systems in the original Illustris simulations were more uniformly distributed around it.
However, all the galaxies (except for system 9, by construction) still fall within the scatter from the SLACS mean relation in the CDM run.  We note that the most massive systems become more extended in the SIDM run, i.e. they have larger effective radii, while their stellar mass only changes by a few per cent.  The right panel of Figure \ref{diagram} shows the same points as the left panel, but color-coded according to the properties of the SIDM density profile, as we will detail in the following paragraph.

\begin{figure*}
\includegraphics[width=0.48\hsize]{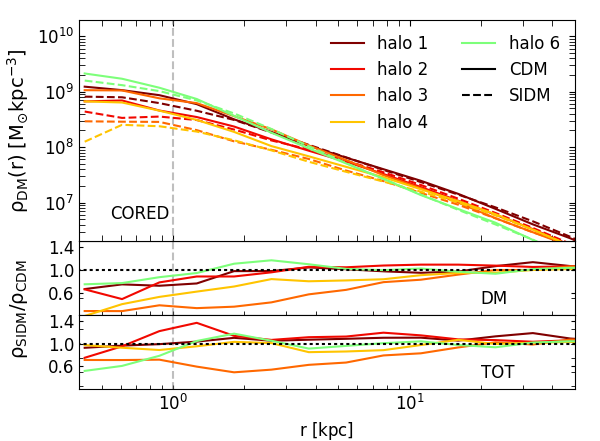}
\includegraphics[width=0.48\hsize]{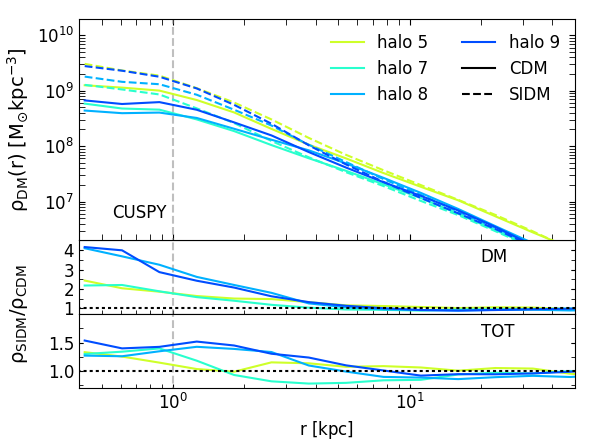}
\caption{Radial density profiles of the main galaxy at $z=0.2$. The CORED systems are shown in the left panels, while the CUSPY ones in the right panels. In the top panels, solid (dashed) lines show the dark matter profile for the CDM (SIDM) run. The ratio between the dark matter and total profiles in the two runs is shown in the middle and bottom residuals panel.  We focus on the inner region of the haloes where the SIDM effects are stronger; the profiles are identical  in the outer parts ($r>0.1\times r_\text{vir}$). The gray vertical like marks the spatial resolution of the simulations, within which interpretation requires care. \label{profile}}

\end{figure*}

We now look at the dark matter and total matter distribution in the SIDM and CDM runs. The top panels of Figure \ref{profile} shows the dark matter radial density profiles of the haloes in both runs: the CDM density is shown by the solid lines of different colours, while the dashed lines stand for the SIDM run. The small middle (lower) panels show the ratio of the SIDM to CDM dark matter (total mass) density profiles. We measure the mean central density  within 15 kpc in both runs (also see Figure \ref{zform} and \ref{profile_evol}) and we use it as a proxy to divide the haloes in two categories: the systems in which the SIDM mean central density is lower than the CDM one are shown in the left panels (hereafter 'CORED' systems), while those in which the central density is higher in SIDM are shown in right panels (hereafter 'CUSPY' systems). Thus, the haloes are in practice classified according to the properties of the SIDM density profile.
It is interesting to note that the difference between the SIDM and CDM density profiles is stronger for the CUSPY systems than for the CORED ones, especially when looking at the dark matter density profile (middle panels). 

The same classification applies to the right panel of Figure \ref{diagram}, where the points in the $r^{*}_\text{eff}-M_{*}$ plane are color-coded according whether they belong to the CORED or CUSPY categories. In general, the green and red squares in the right panel occupy two different parts of the parameter space. The four CUSPY systems occupy the lower-left part of the $r_\text{eff}^{*}-M_{*}$ plane: they have lower stellar masses and, most importantly, their stellar distributions are more compact. A cuspy profile is found in the disk galaxies, but also in the most compact of the ETGs-like systems (halo 5). In this respect, it is of particular interest to compare systems 5 and 6 which, despite having almost identical total and stellar masses in both scenarios, evolve differently in the SIDM run: the stellar effective radius of system 5 is about half of that of system 6, supporting the fact that a very compact baryonic component can dominate the total density profile and drive the growth of a cuspier dark matter density.
This suggests that, apart from the morphological type and the presence of a disk \citep[as in][]{sameie18}, a significant role could be played by the halo and galaxy mass and thus its mass accretion history. We investigate this hypothesis in the next section.

We now fit the dark matter density radial distribution with the Einasto profile \citep{einasto65}, defined as
\begin{equation}
\rho(r) = \rho_{-2}\exp\left\{-2\alpha\left[\left(\frac{r}{r_{-2}}\right)^{1/\alpha} -1\right]\right\},
\end{equation}
where $\rho_{-2}$ and $r_{-2}$ are the density and the radius at which $\rho(r)\propto r^{-2}$ and $\alpha$ defines the steepness of the power-law. The Einasto profile provides a good fit to both the CDM and SIDM haloes, since it is able to well describe different inner slopes. For the CDM haloes, both the NFW and the Einasto profiles provide a good fit, while the NFW profile is not a good fit for cored SIDM haloes; in general, self-interaction produces a larger variety of profile shapes and thus a more flexible definition for the profile is needed. Thus, in what follows we use the Einasto profile for both runs.

\begin{figure}
\includegraphics[width=0.95\hsize]{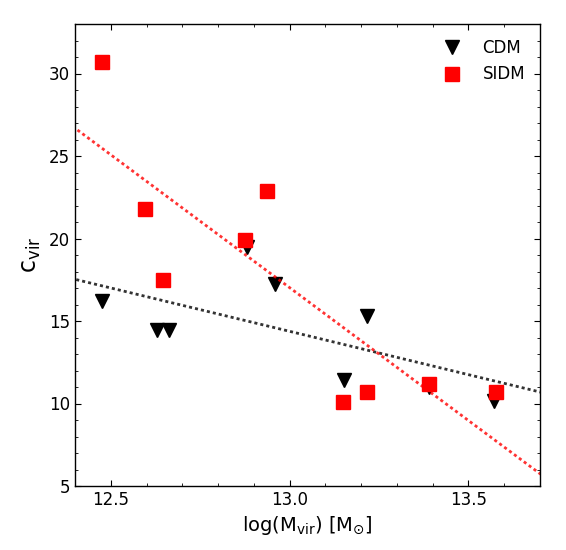}
\caption{Concentration-mass relation at $z=0.2$: $c_\text{vir}$ is calculated from the best fit Einasto profile. The black and red dotted lines show the best-fit linear relation respectively for the CDM and SIDM runs.\label{c-m}}
\end{figure}

Figure \ref{c-m} shows the concentration-mass relation in the two runs: the CDM haloes are represented by the black triangles, while the SIDM ones by the red squares. The dotted lines of corresponding color show the best fit linear relation between $c_\text{vir} = r_\text{vir}/r_{-2}$ and $\log(M_\text{vir})$. The concentrations of our sample are higher than the average concentration-mass relation at $z=0.2$ \citep{duffy08,maccio07}, due to a combination of two effects: $(i)$ the dark matter concentrations in hydrodynamical simulations are altered by the presence of baryons and in particular the IllustrisTNG models produces higher concentrations than in the dark-matter-only run (due to adiabatic contraction of the halo as it responds to the assembly of the galaxy) in the halo mass range spanned by our sample \citep{lovell18} - this effect is enhanced in our case for the CUSPY systems; $(ii)$ we selected analogues of strong lenses, which are in general more concentrated than average at fixed mass \citep{giocoli13}.  
While for CDM the concentration-mass relation is, as usual, a slowly declining function of mass, the SIDM relation is much steeper and thus the haloes show a greater diversity of profiles, in the sense that the SIDM concentrations span a wider range of values than the CDM ones over our limited mass range. This is qualitatively similar to what was found by \citet{robertson18} by looking at SIDM+baryons simulations of cluster-mass haloes. Again, we note how halo 6 follows the behaviour of CORED haloes, while halo 5 that of CUSPY haloes, even though their masses and disk fractions are almost identical in both runs (see Table \ref{sim1}). This suggests that, together with the morphological type and halo mass, something else contributes to shape the final properties of the profile. We investigate this in the next section.

\subsection{Link to the halo evolution} \label{sec_evol}

\begin{figure*}
\includegraphics[width=0.9\hsize]{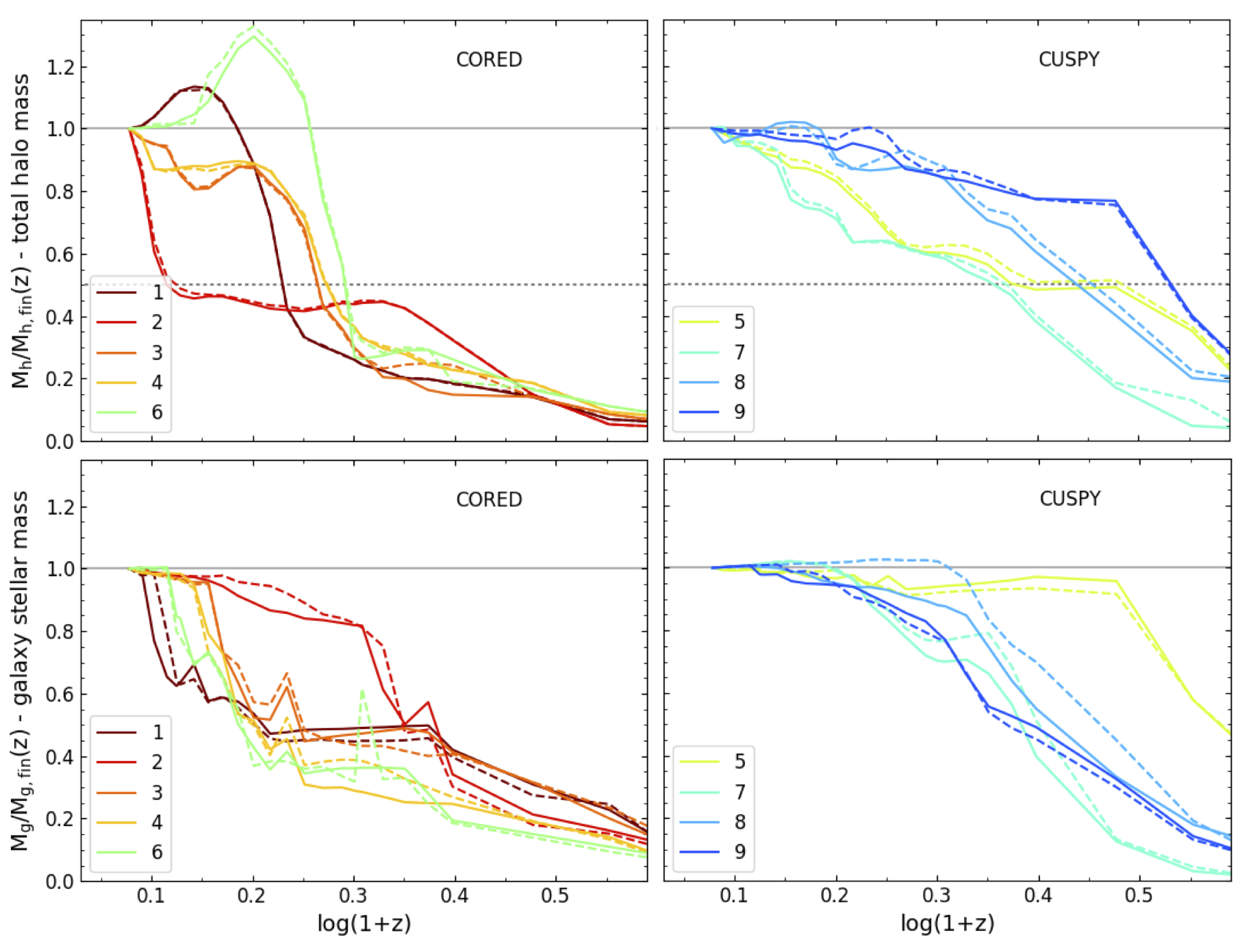}
\caption{Mass accretion history. Results from the CDM and SIDM runs are shown respectively by the solid and dashed lines; the haloes are again divided in CORED (left panel) and CUSPY (right panel) systems, as in Figure \ref{profile}. We show the accretion history of both the halo (total mass within $R_\text{vir}$) and of the stellar mass of the main galaxy. In the top panels, the formation redshift of the haloes ($z_{f}$) correspond to the moment at which the coloured lines intersect the horizontal gray dotted thresholds, corresponding to half of the final halo mass. \label{evol1}}
\end{figure*}

\begin{figure}
\includegraphics[width=0.95\hsize]{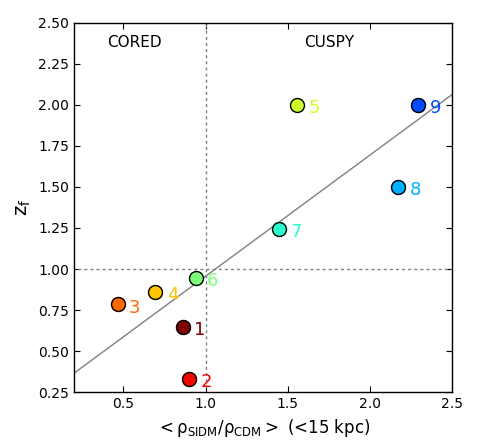}
\caption{Correlation between the properties of the density profile and the halo formation time. The $x$-axis shows the mean ratio of the dark matter density at 1 kpc < r < 15 kpc between the two models, while $z_\text{f}$ is defined as the redshift at which the halo had accreted 50 per cent of its final mass. \label{zform}}
\end{figure}

Figure \ref{evol1} shows the mass accretion history of all the systems from redshift $z=3$ to the final snapshot of the simulation at $z=0.2$, calculated by following the main progenitor back in time. Results from the CDM and SIDM runs are shown respectively by the solid and dashed lines; the haloes are again divided in CORED (left panel) and CUSPY (right panel) systems, as in Figure \ref{profile}. Qualitatively, the two groups of systems behave differently: the CORED systems experience a fast accretion phase at low redshift (late assembly), while the CUSPY systems have a more regular history (early assembly). The accretion history in CDM and SIDM is shown, respectively, by solid and dashed lines: interestingly, in most cases the history of each halo is almost identical in CDM and SIDM, indicating that the mass accretion proceeds similarly (both for DM and baryonic components) since it is regulated by the matter distribution on larger scales, but in the two scenarios it has a different effect on the inner structural properties of haloes. If we define the halo formation time $z_\text{f}$ as the redshift at which the halo mass was already half of the final one in the main progenitor branch \citep{lacey94,giocoli13}, we find that our two categories are also separated in $z_\text{f}$. The formation redshift $z_\text{f}$ of each halo is the same in both runs, due to the similarity in the mass accretion history. The value of $z_{f}$ correlates with the differences in the dm-profile between the two runs: Figure \ref{zform} shows $z_{f}$ as a function of the mean ratio of the central density, measured within 15 kpc from the centre. The coloured points show the position of each system and the solid grey line is the best-fit linear relation between the two quantities, which represents well the general trend. The two dotted lines separate the regions of the parameter space into $(i)$ CORED vs. CUSPY (vertical) and $(ii)$ $z_\text{f}<1$ vs $z_\text{f}>1$ (horizontal). 

\begin{figure*}
\includegraphics[height=0.38\hsize]{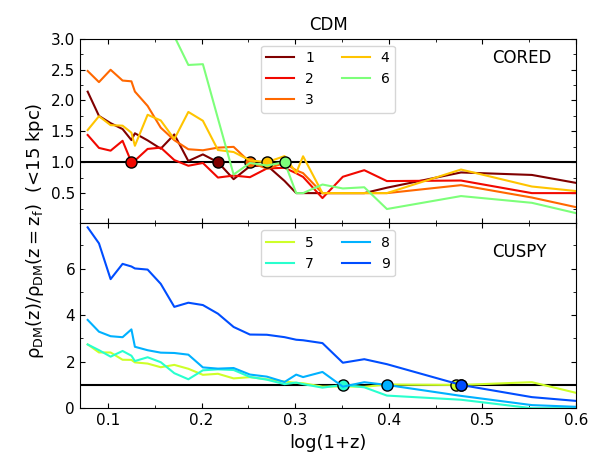}
\includegraphics[height=0.38\hsize]{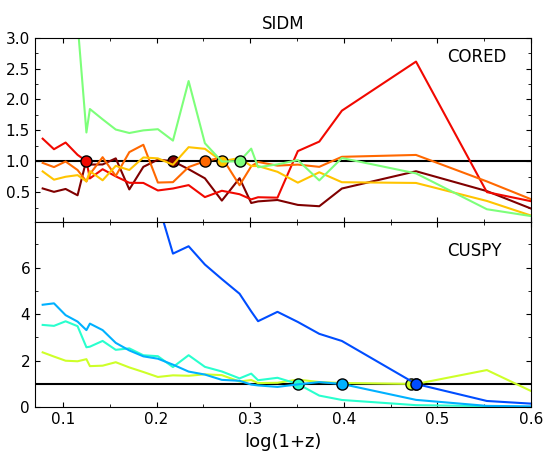}
\caption{ Evolution with redshift of the mean central density in dark matter in the CDM (left) and SIDM (right) scenarios. Each coloured line stands for one of the haloes in the sample and the points of corresponding colour mark $z=z_\text{f}$. While for CDM the central density steadily increases with time, in SIDM this is true only for the CUSPY haloes. \label{profile_evol2}}
\end{figure*}

Figure \ref{profile_evol2} shows the evolution with redshift of the mean central density in dark matter of all the sample in the CDM (left panels) and SIDM (right panels) scenarios. The $y$-axis shows the mean value of $\rho_\text{DM}(z)/\rho_\text{DM}(z=z_\text{f})$, measured in the inner 15 kpc from the centre (1 < r < 15 kpc) as in Figure \ref{zform}.  The formation redshift $z_{f}$ is marked by a coloured circle for each halo. While for CDM the central density steadily increases with time, in SIDM this is true only for the CUSPY haloes; the central density in most of the CORED systems oscillates around $\rho_\text{DM}(z=z_\text{f})$ with an irregular and more complex behaviour, in particular at $z<z_{f}$.

\begin{figure*}

\includegraphics[width=0.95\hsize]{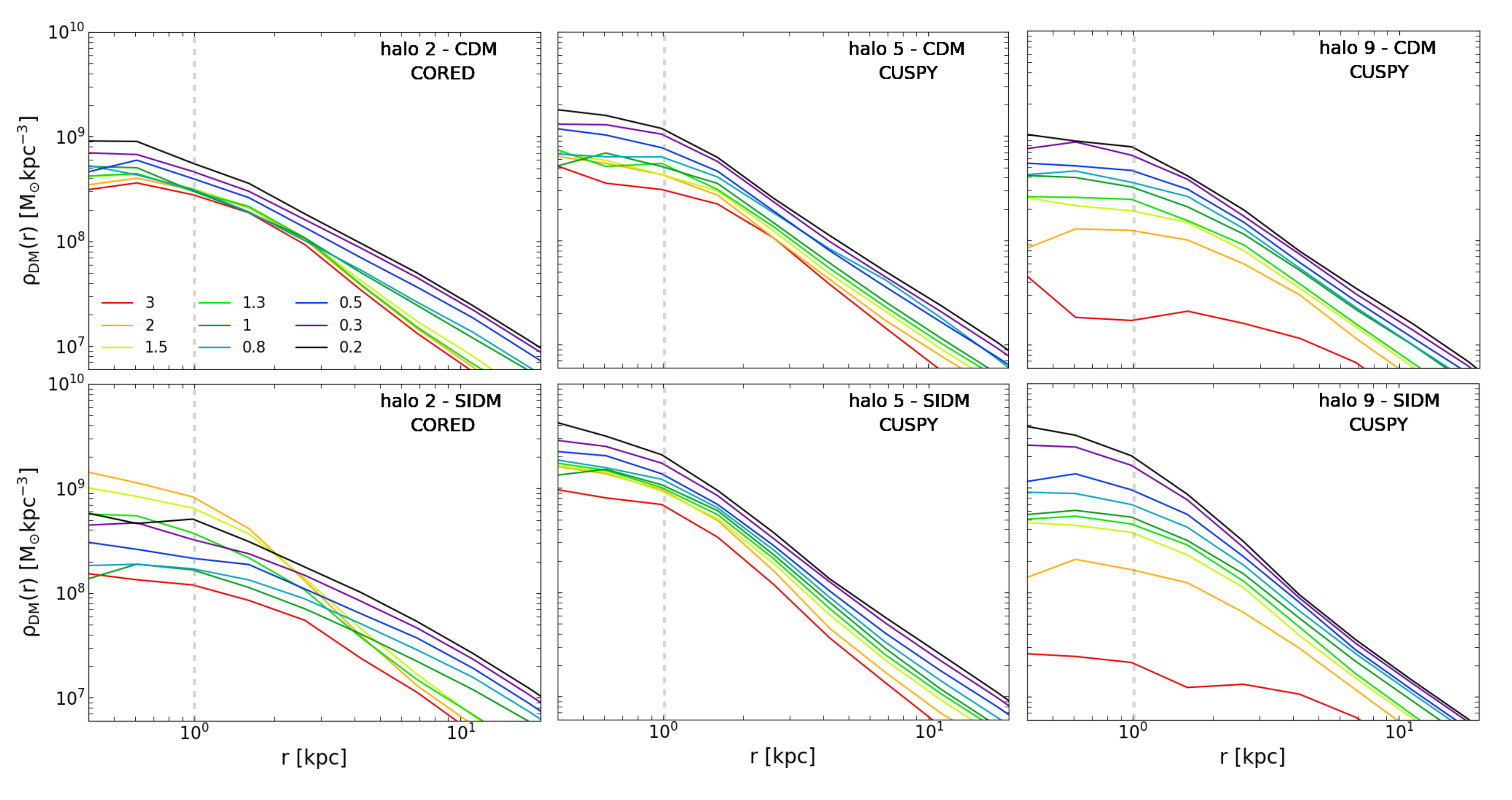}
\caption{Evolution with redshift of the inner part of the density profile (1 kpc < r < 15 kpc) with redshift in the CDM (first row) and SIDM (second row) runs. We show three haloes, as an example of the three main behaviours.  The gray vertical like marks the spatial resolution of the simulations, within which interpretation requires care. \label{profile_evol}}
\end{figure*}

Our sample is not large enough to fully characterize the halo evolution, but we believe that it spans a range in mass and morphology that is wide enough to derive a consistent picture. The correlation between the mass accretion history and the properties of the final density profile suggests that in the CUSPY systems the baryons drive the evolution of the central dark matter core - created at an earlier time by the DM self-interaction (since haloes are assembled earlier) - and the interplay between baryons and dark matter leads to the final cuspy profile. The more massive haloes, belonging to the CORED category, formed at redshift $z\leq 1$ and thus we see them at the stage in which the large central core is just established. Moreover, a very compact stellar component ( and the morphology of the central galaxy, i.e. the presence of a stellar disk) enhances the different behaviours, contributing to the formation of an even cuspier profile, while this is not the case for the more extended baryonic distribution of the ETGs (see the effective radii in Figure \ref{diagram}).
In order to better visualize these behaviours, in Figure \ref{profile_evol} we show the dark matter density profile of three haloes from our sample, from $z=3$ to $z=0.2$. The top row shows the evolution in the CDM run, which is qualitatively similar for all the haloes in our sample: the main progenitor dark matter central density increases with time. The second row shows the corresponding profiles in the SIDM run. In this case, there is a larger halo-to-halo variation. 
the SIDM dark matter density profile of Halo 2 (CORED) clearly oscillates with time, going through an initial growth phase ($z>2$) during which the stellar distribution becomes more compact and the central density increases, followed by the creation of a central core ($0.8\leq z\leq 2$) and by a final phase in which the central density rises again. In the other CORED systems, the oscillations of the central density are less pronounced, but consistent with the same behavior.
On the other hand, Halo 9 shows one of the cuspiest SIDM profiles in our sample and in this case, the SIDM dark matter profile grows monotonically as in CDM and reaches a higher central density, suggesting that the presence of a disk galaxy triggers an evolution similar to the one seen in \citet{sameie18}. Nevertheless, in the same work it is also claimed that compact stellar disks have a stronger effect than extended ones. Our simulations seem to suggest once again that, in the presence of a realistic baryonic physics model, the co-evolution of the two components is more complex: as can be seen from Table \ref{sim1}, the central galaxy of Halo 9 has a more massive and extended disk with respect to Systems 7 and 8, but the total variation of the density profile is similar.
Finally, Halo 5 (middle column) belongs to the CUSPY category at $z=0.2$, but has an intermediate total and stellar mass. In this case, the evolution of the density profile is also somehow in between the two categories since the final SIDM profile has a slightly cuspier final profile and a higher central density, but not as extreme as Halo 9.

\subsection{Lensing properties} \label{sec_lensing}

Since gravitational lensing allows to measure the total projected mass distribution of the deflectors directly, the results presented in the previous sections may have direct consequences on the distribution of the main lensing observables, such as the Einstein radii $r_{E}$.

We use the lensing code {\sc GLAMER} \citep{metcalf14,petkova14} to create realistic mock images from the simulations. Using the particle positions and masses, {\sc GLAMER} calculates the projected surface mass density and thus the convergence, magnification and deflection angles created by the main lens with a tree algorithm. Each particle is represented with a B-spline in three dimensions as is commonly done in smooth particle hydrodynamics (SPH) simulations. The dark matter particles in simulations are point-like, but in order to avoid unrealistic deflections of the light rays, their mass has to be `smoothed', i.e. distributed on a larger physical scale, according to their surrounding density so that the halo appears as a continuous mass distribution. The size of the smoothing length is set to the distance to the $N_{\rm smooth}$-th nearest neighbor where ${N_\text{smooth}}$ can be adjusted - in this work $N_\text{smooth}=64$.  This scheme provides smaller smoothing lengths where the particles are dense (for example at the centers of haloes) and it is justified by a large number of particles, and larger ones where the particles are sparse and shot noise would otherwise be a problem. A similar approach is followed also for the baryonic cells, since {\sc GLAMER} is not able to take into account the shape of each individual cell. 

\begin{figure}
\includegraphics[height=0.95\hsize]{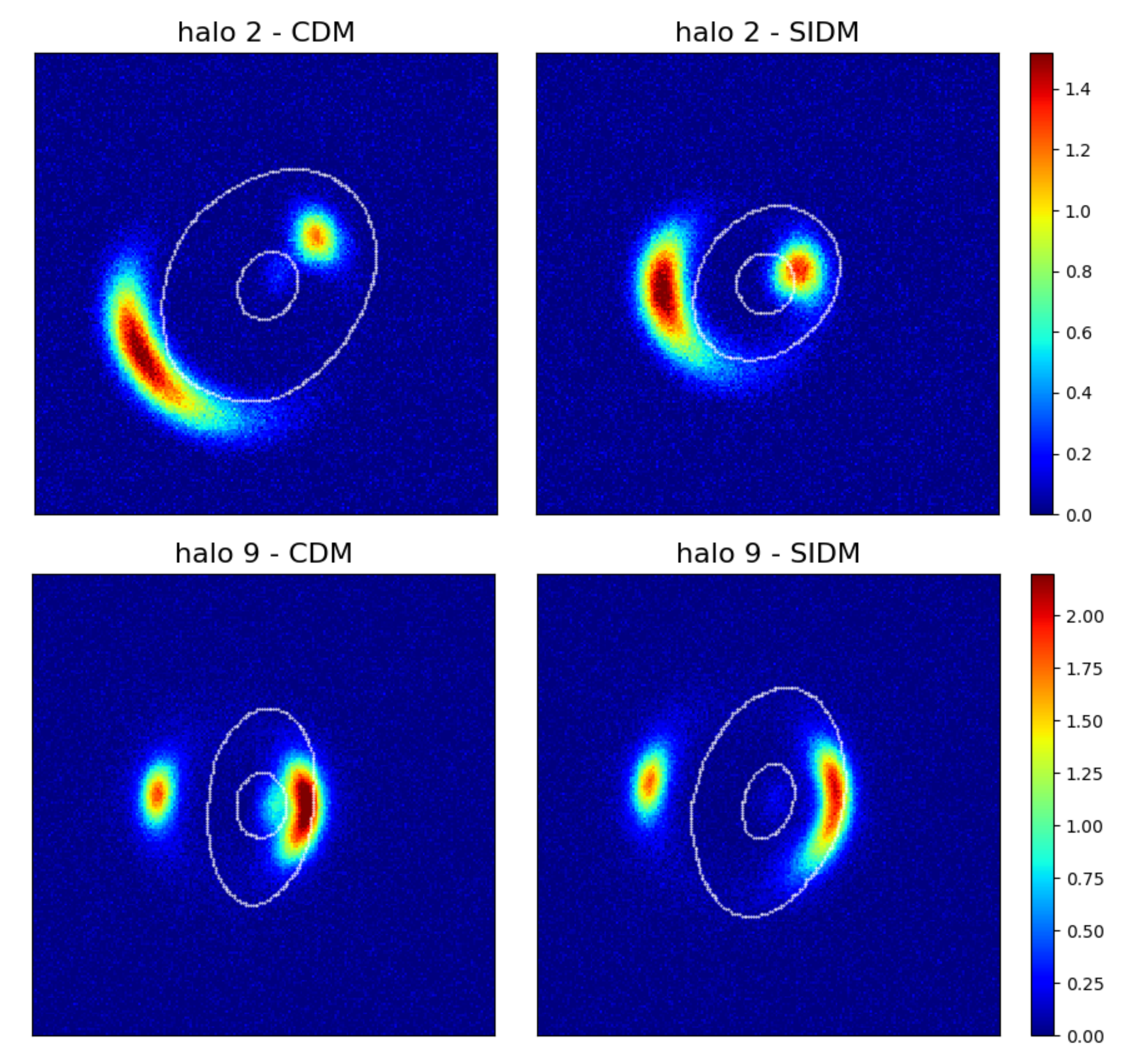}
\caption{Mock lensed images created from two of the simulated haloes, for a simulated source at $z=1$. The source is a mock image of an early-type galaxy from the Illustris simulation, created in the ACS-F606 filter, using the technique described in \citep{torrey15}; we used the same source, located in the same position with respect to the center of the image plane, for the four cases. It has then been convolved with a HST-like point spread function and a realistic Gaussian noise has been added. The critical lines are shown in white; all the images are 5 arcsec on a side. \label{images}}
\end{figure}

We ran the ray-tracing code through 200 different random projections for each halo to have a good statistical sample; the lens and source redshifts are set respectively to $z_{L}=0.2$ and $z_{S}=1$.  Figure \ref{images} shows examples of the mock images created with a simulated background source, for two haloes in the sample; the images are 5 arcsec wide and the critical lines (in white) are superposed to the surface brightness distribution of the lensed images. 

\begin{figure*}
\includegraphics[width=0.9\hsize]{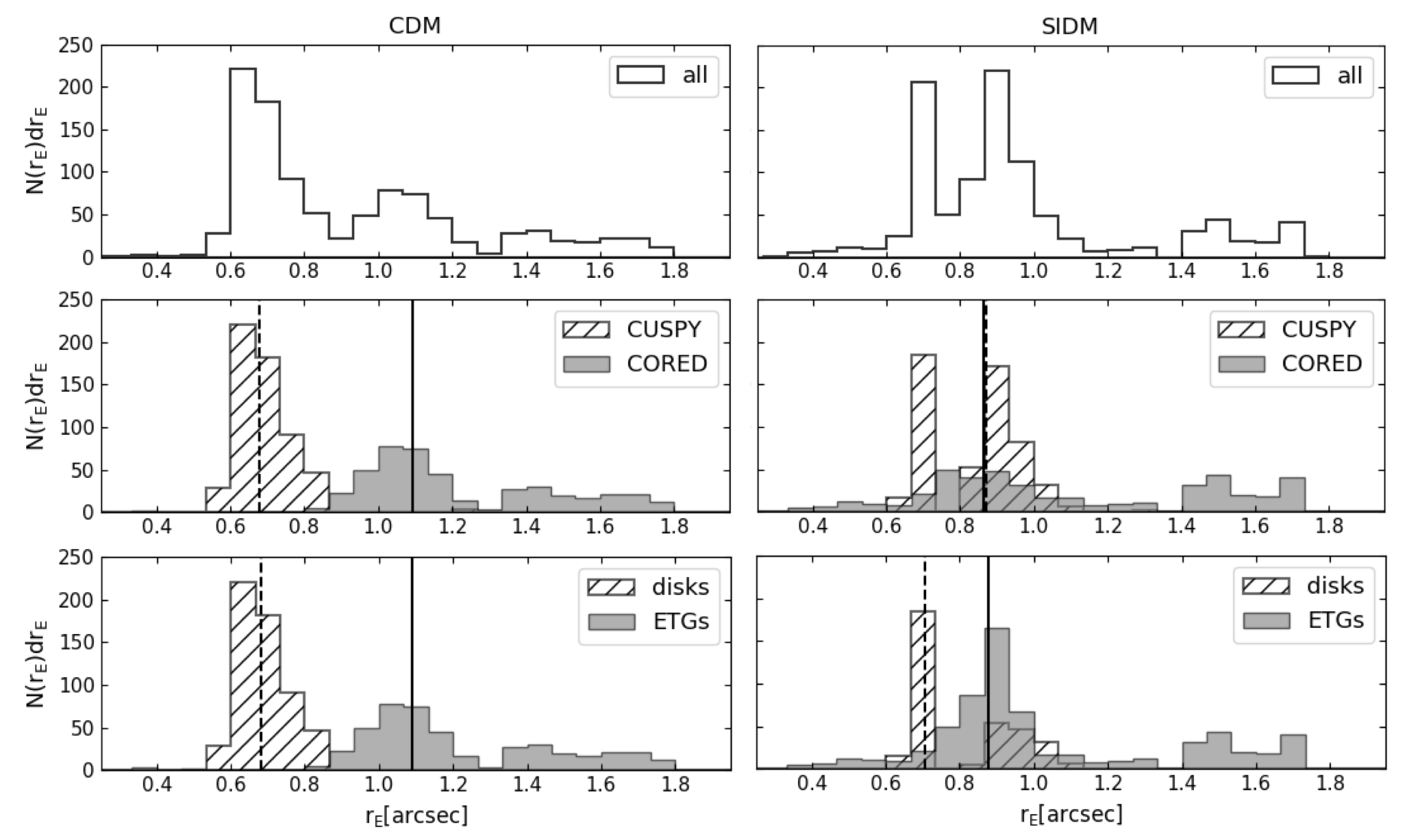}
\caption{ Einstein radii distribution for the ray-tracing runs in the CDM (left) and SIM (right) scenarios. The top panels shows the overall distribution derived from 200 random projections for each halo, while in the second and third rows galaxies are divided according to the properties of the corresponding SIDM density profile or their morphological type in each run (based on the D/T ratio). The vertical solid and dashed lines show the medians for the two populations.  \label{ratio_rad}}
\end{figure*}

From each of the mock images, we estimated the Einstein radius $r_{E}$, whose size provides a way to quantify the magnitude of the lensing power of the halo. In the top panels of Figure \ref{ratio_rad}, we show the distribution of the Einstein radii for the CDM (left) and SIDM (right) runs from the whole sample of mock images (i.e. 200 projections for each of the nine haloes). Then, we separate the haloes in two categories according to the properties of their SIDM density profile (middle) or the morphological classification of the main galaxy (bottom), using the D/T fractions reported in Table \ref{sim1}. In the CDM run, the last three haloes are classified as disks, while in SIDM halo 8 has a much smaller disk component and thus belong to the other category. The median Einstein radii of the two distributions are shown by the dashed (for CUSPY or disks) and solid (for CORED or ETGs) vertical lines.
The Einstein radii of ETGs peak at lower values in SIDM than in CDM, due to their cored density profiles. In particular, due to the central density cores in ETGs, the SIDM distribution lacks the largest Einstein radii. It is also interesting to note that in SIDM the median Einstein radius of CORED and CUSPY systems are in practice identical and that the two distributions superpose: the bi-modality seen in CDM is in general lost in the SIDM run. These results seem to suggest that the observed distribution of Einstein radii in a large survey such as EUCLID could in principle allow us to distinguish between SIDM and CDM and provide an independent probe of the nature of dark matter. In practice, a wider distribution of source and lens redshifts and a wider sample of simulated galaxies resembling that predicted for EUCLID or other wide-field surveys should be used to test whether our current results hold in a realistic scenario.

In this respect, it is also interesting to note that some properties of systems 3 and 6 do not match the ones of SLACS lenses: even if they can be classified as early-types, their sizes are larger than the SLACS ones and they lie farther away from the mean relation shown in Figure \ref{diagram}, which in turn corresponds to the fact that their Einstein radii are smaller in SIDM than in CDM. Thus, had the selection been done directly on the SIDM run, they might not have been chosen as SLACS analogues.

\section{Subhalo population} \label{sec_sub}

In this section, we study the properties of the subhaloes in our sample. In particular, we measure the subhalo mass function in the two runs and we compute the subhalo density profiles. SIDM models are often mentioned alongside warm dark matter as able to resolve a number of discrepancies between observations and numerical simulations. Moreover, the presence of the self-interaction can alter the number density and the properties of subhaloes, and with them their detectability through gravitational lensing \citep{vegetti10,vegetti12,vegetti14,hezaveh16,li16b,despali18,birrer16}. 

\begin{figure}
\includegraphics[width=\hsize]{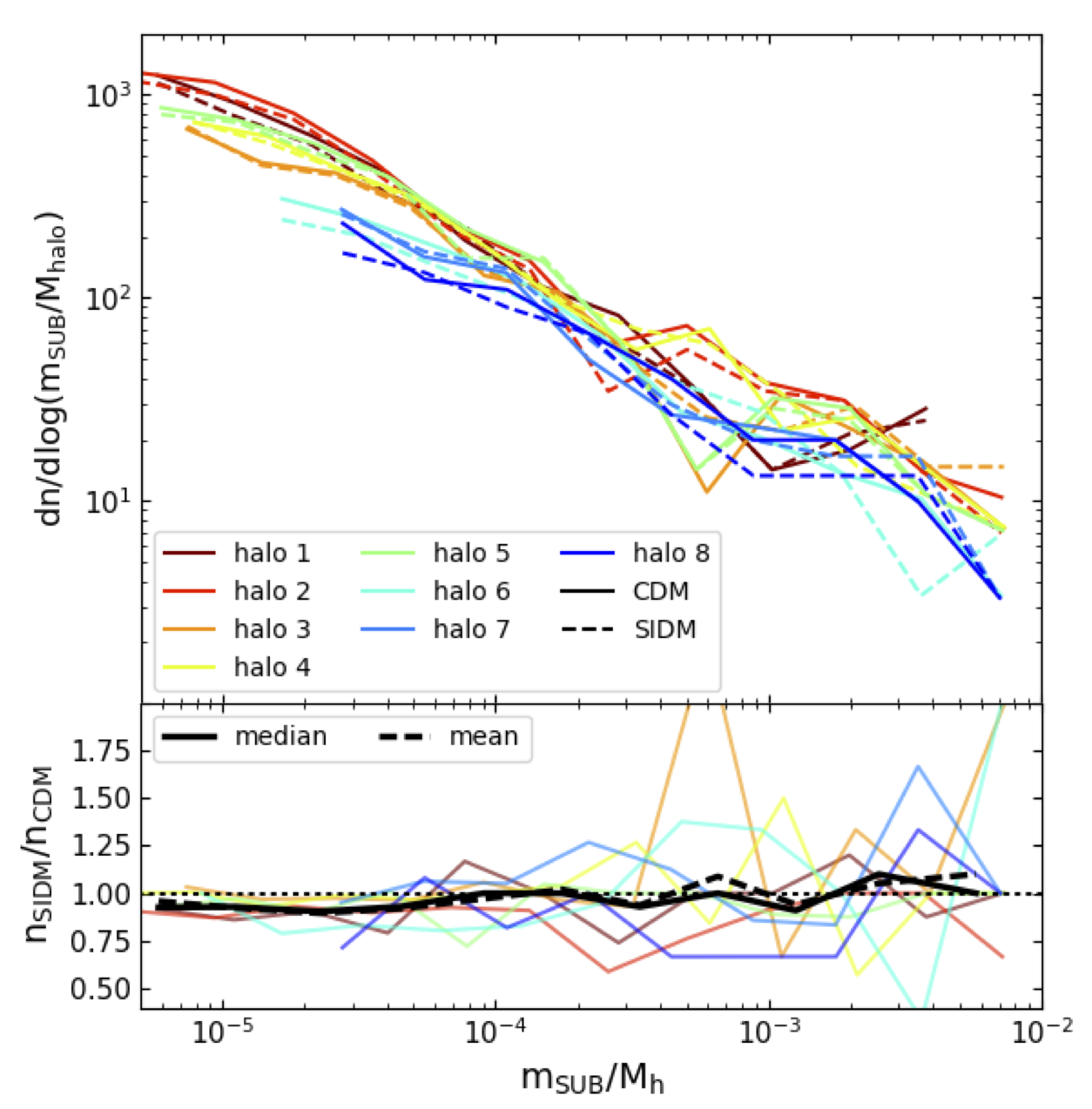}
\caption{Subhalo mass function. In the top panel we show the subhalo mass function of the simulated haloes, both in the CDM (solid lines) and SIDM scenarios (dashed lines). The ratio between the number counts in the two runs is shown in the bottom panel: on average, there is no significant difference in the subhalo number counts, as can be seen from the mean ratio (solid black line)  \label{submf}}
\end{figure}

 In the top panel of Figure \ref{submf} we show the subhalo mass function for all the subhaloes with a mass $M_\text{sub}>10^{8}M_{\odot}$, while in the bottom panel we plot the ratio of the number counts in the two runs for each halo and the mean and median ratios (solid and dashed black lines). For the SIDM model used in this work, we find no significant difference in the subhalo counts with respect to the CDM scenario. This result is in agreement with previous findings from \citet{zavala13} and \citet{rocha13}. In particular, \citet{zavala13} have found that a much higher cross-section (10 $\text{cm}^{2}\text{g}^{-1}$) is necessary to have a significant suppression in the number of subhaloes; at the same time, they found that a much lower cross section of 0.1 $\text{cm}^{2}\text{g}^{-1}$ is in this context indistinguishable from CDM. 
 Recently, \citet{rivero18b} have measured the subhalo power spectrum in CDM and the ETHOS model \citep{vogel16} from maps of projected mass density (convergence), finding a clear suppression and thus suggesting that gravitational lensing might be used to distinguish the two scenarios. However, it should be pointed out that this result is not related to the presence of self-interaction (too weak to suppress the number density in their model), but mainly to a truncation of the initial power-spectrum at the small scales. Thus, while in general velocity independent SIDM models with a low cross-section do not affect the number of subhaloes in an appreciable way,  it may be that more complicated models that include other effects other than self-interaction - such as velocity independent cross sections, inelastic scattering or dark radiation \citep{vogel16,vogel18b} - may result in significant changes. Moreover, models including a truncated power-spectrum may resemble WDM from the point of view of subhalo abundances.

\begin{figure}
\includegraphics[width=\hsize]{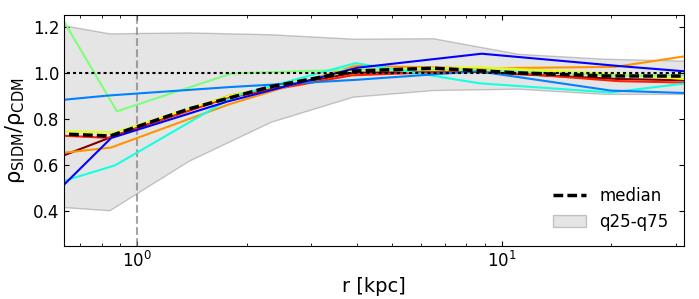}
\caption{Ratio of the subhalo density profiles, for systems with $N_\text{part}>500$. The median ratio for each halo is shown by a different color, while the median ratio for the whole sample  is represented by the black dashed line. The gray shaded area encloses the region between the 0.25 and 0.75 quantiles. The gray vertical line marks the spatial resolution of the simulation, within which interpretation requires care. \label{sub_prof}}
\end{figure}

In order to directly compare subhalo properties between the two runs, we selected the matching subhaloes using the particles IDs: if the matched subhaloes share more than 50 per cent of the mass, we consider the matching to be reliable. The masses of the matched systems are on average the same in the two runs, with a maximum difference of about 15 - 20 per cent. We measured the total density profiles of each pair and then checked whether the slope of the profile is different: Figure \ref{sub_prof} shows the median ratio of the subhalo density profiles of matched subhaloes. We use only the subhaloes with more than 500 particles, corresponding to $M_\text{sub}\geq1.4\times 10^9 M_{\odot}$, in order to have a reliable estimate of the density profile. The median central density of subhaloes is at most 20 per cent lower in the SIDM scenario (black dashed line) for the region in which the profile can be trusted. However, the gray shaded region marks the region enclosed between the 0.25 and 0.75 quantiles, still consistent with no differences between the two profiles. Given that our sample is limited, we can only conclude that we expect little difference in the lensing signal caused by subhaloes and thus on their detectability through lensing, between this model and CDM. However, other SIDM models may cause stronger effects on the subhalo profiles (as on their abundances), creating a further degeneracy with WDM models. Detailed studies including different SIDM models are necessary to fully understand how well SIDM, WDM and CDM can be discriminated on the basis of the subhalo number density and of the corresponding lensing signal, in particular. 

\section{Discussion} \label{discussion}

In this work, we have analysed a sample of simulated galaxies in CDM and SIDM scenarios, in order to understand the effects of dark matter self-interaction on the final halo properties. For this purpose, we selected nine galaxies from the Illustris \citep{vogel14} simulation and re-simulated them in CDM and SIDM with the IllustrisTNG model for baryonic physics. Our sample consists of six early-type galaxies (analogues of observed systems, such as the SLACS lenses) and three galaxies with a fairly important disk component (see Table \ref{sim1}). 
The main result of this work is that, in the presence of both dark matter self-interactions and baryonic physics, halo density profiles have a complex evolution and a larger variety of final properties than dark-matter-only SIDM simulations (or CDM simulations). Our analysis is generally consistent with the results from \citet{sameie18} and \citet{robertson18}, but using high-resolution hydrodynamical simulations allowed us to reach a deeper understanding of the halo evolutionary properties.

While in dark-matter-only SIDM simulations all haloes create a central core, with a size proportional to the strength of the self-interaction \citep{vogel14b}, in the presence of baryonic physics only part of our sample shows a cored profile in the SIDM run (five out of nine simulated systems); four systems, on the contrary, have a cuspier profile in SIDM than in CDM (Figure \ref{diagram} and \ref{profile}). 

We divide the haloes in two categories (CORED and CUSPY), according to whether the final SIDM dark-matter density profile is flatter or steeper than the CDM one. We find that the final properties, such as the density profile, correlate with the merging history of the halo: CORED systems are in general more massive and formed at $z\leq 1$, showing a recent phase of fast accretion, while CUSPY systems formed at $z\geq 1$ (Figure \ref{evol1}). The dark-matter density profiles also show a more complex evolution with time in the SIDM scenario: the central density oscillates in CORED systems, where the self-interaction acted recently to create a core, while the combination of early formation time and the presence of a disk increases the central density in CUSPY systems. A general consequence is that SIDM haloes possess a greater variety of density profiles and concentrations with respect to CDM.

The size of our sample is limited and thus we cannot draw statistically-supported conclusions. Nevertheless, the systems span a range in halo properties (such as mass and morphological type) that is sufficient to retrieve a physically-consistent picture. We confirm the conclusion from previous works \citep{kaplinghat14,sameie18,robertson18}, that the baryonic physics has a stronger impact on the final density profile in SIDM models: this is demonstrated by the fact that the haloes can be efficiently classified in two separate categories based on their final properties. However, these final properties also correlate with the mass accretion history: systems with very similar mass and morphological types can belong to different categories and their behaviour can be understood in terms of the difference in the accretion history.
This correlation suggests that in the CUSPY systems the self-interaction acted at earlier time, but then the baryons dominate the evolution of the density profile and the interplay between baryons and dark matter leads to the final cuspy profile. This effect is enhanced in the presence of a compact baryonic distribution (see the effective radii of CUSPY and CORED systems in Figure \ref{diagram}) and the presence of a stellar disk, which drive an extra boost in the central dark matter density profile in SIDM simulations, with respect to systems with an ETG-like morphology.
On the other hand, the more massive haloes (CORED) formed at redshift $z\leq 1$ and thus we see them at the stage in which the large central core is established.
As a consequence of this complex interplay between SIDM and baryonic physics, also the lensing properties of our sample are different in the two runs. First of all, due to the presence of the central core, some of the most massive systems would not be classified as SLACS lens analogues (see Figure \ref{diagram}) in SIDM. Moreover, the overall distribution of Einstein radii is different in the two scenarios. In CDM the overall distribution is clearly bi-modal and the contributions of ETGs-analogues and disk galaxies are separated, while in SIDM the mean Einstein radii of these two categories are closer to each other and the two distribution partially superpose. In particular, due to the central density cores in ETGs, the distribution lacks the largest Einstein radii. Our results suggest that future large surveys could provide a way to constrain the dark matter model and exclude some scenarios based on the overall distribution of lensing properties, even though a larger sample and a wider distribution of lens and source redshifts needs to be considered to confirm this result. 
We remark that different results might be obtained by using a reduced number of zoom simulations or a full cosmological simulation, as shown for instance in the discrepancies in the properties of SIDM haloes in the cluster mass range between the results of \citet{robertson18} and \citet{robertson18b}. Thus, our results should be extended to a full cosmological box in order to be confirmed.

Finally, we have analysed the subhalo population in the two runs. We find that the number density of subhaloes is not significantly different in the two considered scenarios: from a lensing point of view, this means that CDM and our adopted SIDM model could not be clearly distinguished with techniques such a gravitational imaging \citep{vegetti10,vegetti14}.

Our results clearly show that it is of fundamental importance to take into account the effects of baryonic physics, especially when comparing to observational results. In the future, the interplay between SIDM models and baryonic physics should be investigated using larger halo samples or cosmological simulations - and more SIDM models (for example with a velocity-dependent cross section, ETHOS or inelastic models), in order to study its effect on a broader mass range.

\section*{Acknowledgements} 
We thank the IllustrisTNG collaboration for allowing the use of the new IllustrisTNG model for this work. We thank Annalisa Pillepich and Carlo Giocoli for their useful comments and Ben Metcalf for giving us access to the lensing code {\sc GLAMER}.
MS acknowledges support by the European Research Council under ERC-CoG grant CRAGSMAN-646955. SV has received funding from the European Research Council (ERC) under the European Union's Horizon 2020 research and innovation programme (grant agreement No 758853). MV acknowledges support through an MIT RSC award, a Kavli Research Investment Fund, NASA ATP grant NNX17AG29G, and NSF grants AST-1814053 and AST-1814259. JZ acknowledges support by a Grant of Excellence from the Icelandic Research Fund (grant number 173929$-$051).  This research made use of Astropy, a community-developed core Python package for Astronomy \citep{astropy13,astropy18}.

\bibliographystyle{mnras}
\bibliography{mnras_template.bbl}
\label{lastpage}
\end{document}